\documentclass[pre,twocolumn,showpacs,floatfix]{revtex4}
 \usepackage{graphicx}
 \newcommand{\be}{\begin{equation}}
 \newcommand{\ee}{\end{equation}}
 \newcommand{\bea}{\begin{eqnarray}}
 \newcommand{\eea}{\end{eqnarray}}

\begin{document}

 \title{Nonequilibrium phase transitions in models of adsorption and
desorption}

 \author{R. Rajesh}
 \email{rrajesh@brandeis.edu}
 \affiliation{Martin Fisher School of Physics, Brandeis University,
 Mailstop 057, Waltham, MA 02454-9110, USA}
 \date{\today}

 \begin{abstract}
 The nonequilibrium phase transition in a system of diffusing,
coagulating particles in the presence of a steady input and evaporation
of particles is studied. The system undergoes a transition from a phase
in which the average number of particles is finite to one in which it
grows linearly in time. The exponents characterizing the mass
distribution near the critical point are calculated in all dimensions. 
 \end{abstract}
 \maketitle

\section{\label{sec1} Introduction}

There is a variety of physical phenomena in which the processes of
diffusion, coagulation, adsorption and desorption play an important
role.  For example, submonolayer epitaxial thin film growth involves
deposition of atoms onto a substrate and diffusion of these atoms
leading to their aggregation into islands of increasing size
\cite{epitaxy}. A second example is river networks which have been
modeled by aggregating masses in a steady influx of particles
\cite{scheidegger,dodds}.  Further examples include aerosols and clouds
\cite{fried}, colloids \cite{white}, and polymerization \cite{ziff}.

A simple lattice model that incorporates the above processes is the
In-Out model \cite{inout1,inout2} in which diffusing point size
particles on a lattice coagulate together on contact forming particles
of larger mass.  In addition unit mass is input uniformly at rate $q$
while unit mass evaporates from an existing mass at rate $p$.  The
competition between adsorption and desorption results in a
nonequilibrium phase transition between a phase in which the average
mass in the system is finite to one in which it increases linearly with
time. A quantity that captures the features of the steady state is the
mass distribution $P(m,t)$. $P(m,t)$ is the probability that a randomly
chosen site has mass $m$ at time $t$.  For fixed desorption rate $p$,
the distribution $P(m,t)$ for large times was shown to change from an
exponential distribution at small values of $q$ to a power law
$m^{-\tau_c}$ at $q=q_c(p)$ and to a different power law $m^{-\tau}$ for
$q>q_c(p)$. Near the transition point, $P(m,q-q_c,t)$ was seen to have
the scaling form $P(m,q-q_c,t) \sim m^{-\tau_c} Y\left[m (q-q_c)^\phi, m
t^{-\alpha}\right]$. The values of these exponents in high dimensions
were calculated using a mean field approximation. In one dimension, they
were determined using monte carlo simulations. The numerical values in
one dimension were significantly different from the mean field results.
In this paper, these exponents are calculated in all dimensions. The
model is also extended to one in which particles diffuse with a mass
dependent rate $m^{-\mu}$ with $\mu \geq 0$. The critical exponents for
this more general model are also calculated.

Related models have been studied in the context of nonequilibrium
wetting \cite{alon,politi}.  If the mass $m$ in the In-Out model is
identified with the height of a substrate, then the transition observed
is qualitatively similar to nonequilibrium wetting transitions. In these
models, the system undergoes a transition from a phase in which the
interface is smooth to one in which it is rough. The exponents
describing these transitions have been found to be related to some
underlying contact process undergoing an absorbing to active transition.
The In-Out model studied in this paper differs from these models by the
lack of a surface tension term which tries to smoothen out the
interface, and thus belongs to a different universality class. 

The rest of the paper is organized as follows. In Sec.~\ref{sec2}, the
model is defined, the results obtained in Ref.~\cite{inout2} for the
$\mu=0$ case are reviewed, and the results of this paper are summarized.
 In Sec.~\ref{sec3} a scaling relation is derived between the critical
exponents. In Sec.~\ref{sec4}, the phase in which the mean mass
increases linearly with time is fully characterized. In Sec.~\ref{sec5},
exactly solvable limits of the model, namely $\mu=1$ case, zero
dimensions and mean field solution, are discussed.  In Sec.~\ref{sec6},
the exponents for arbitrary $\mu$ are derived in all dimensions with the
help of an assumption. The results are compared with monte carlo
simulations in one dimension. In Sec.~\ref{sec7}, the results of the
model are compared with results of related models.  Section~\ref{sec8}
contains a summary and concluding remarks.  The appendices contain the
details of the calculations.

\section{\label{sec2} Model and Results}

\subsection{\label{sec2a} Definition}

For simplicity, we define the model on a one dimensional lattice with
periodic boundary conditions; generalizations to higher dimensions is
straight forward. Each site $i$ of the lattice has a non-negative
integer mass variable $m_i \geq 0$.  Given a certain configuration of
masses at time $t$, the system evolves in an infinitesimal time $dt$ as
follows. A site $i$ is chosen at random (with probability $dt$), and then 
the following events
can occur.  (i) Adsorption: with probability $q/(p+q+1)$, unit mass is
adsorbed at site $i$; thus $m_i \rightarrow m_i+1$.  (ii) Desorption: if
the mass $m_i$ is greater than zero, then with probability $p/(p+q+1)$,
unit mass is desorbed from site $i$; thus $m_i \rightarrow m_i -1$
provided $m_i \geq 0$.  (iii) Diffusion and aggregation: if the mass
$m_i$ is greater than zero, then with probability $m_i^{-\mu}/(p+q+1)$
the mass $m_i$ moves to a randomly chosen nearest neighbor.  If the
target site already happens to have some mass, then the total mass just
adds up; thus $m_i \rightarrow 0$ and $m_{i\pm 1} \rightarrow m_{i\pm 1}
+ m_i$.  The initial condition is chosen to be to be one in which all
sites have mass zero. The model has three parameters, $p,q,\mu$. 

\subsection{\label{sec2b} Review of results for $\mu=0$}

For $\mu=0$, when all particles diffuse at the same rate, the single
site mass distribution at time $t$, $P(m,t)$ was determined in large
dimensions using a mean field approximation and in one dimension using
monte carlo simulations \cite{inout2}. It was shown that when the
adsorption rate $q$ was increased keeping the desorption rate $p$ fixed,
the system undergoes a nonequilibrium phase transition across a critical
line $q_c(p)$ from a phase in which $P(m)= \lim_{t\rightarrow \infty}
P(m,t)$ has an exponential tail to one in which it has an algebraic tail
for large mass; i.e,
 \be
 P(m) \sim \cases{
 e^{-m/m^*} & when $q < q_c$, \cr
 m^{-\tau_c} & when $q = q_c$, \cr
 m^{-\tau} & when $q > q_c$, \cr}
 \label{eq:1}
 \ee
 where $m^*$ is a $q$ dependent cutoff, and $\tau$ and $\tau_c$ are
exponents characterizing the power law decay. In addition, it was argued
that as a function of the small deviation $\tilde{q} = q-q_c$, and large
time $t$, $P(m, \tilde{q}, t)$ displays the scaling form
 \be
 P(m,\tilde{q}, t) \sim \frac{1}{m^{\tau_c}} Y \left( m \tilde{q}^\phi,
\frac{m}{t^\alpha} \right),
 \label{eq:2}
 \ee
 in terms of three unknown exponents $\phi$, $\alpha$, $\tau_c$, and the
two variable scaling function $Y$. The the three phases will be called
as the {\it exponential} phase ($q<q_c$), the {\it critical phase}
($q=q_c$) and the {\it growing} phase ($q>q_c$). 

Of interest are two more exponents. The mass at each site could be
thought of as representing the height of an interface. In that case, two
quantities of interest are the average velocity of the interface
$F=d\langle m \rangle /dt$ and the fluctuations of the interface about
its mean which would be dominated by $\langle m^2 \rangle$, where
$\langle \ldots \rangle$ denotes a spatial average. In the exponential
phase, $\langle m \rangle$ is finite and hence $F=0$. In the growing
phase, the interface has a finite velocity, and the velocity increases
from zero as $F \sim \tilde{q}^\theta$, where $\theta$ is an exponent.
At the critical point $\langle m^2 \rangle \sim t^\beta$. Using the
scaling form Eq.~(\ref{eq:2}), it is straightforward to derive
\cite{inout2}
 \bea
 \theta & = &\phi \left[ 1- \alpha (2-\tau_c) \right] /\alpha,
\label{eq:3}\\
 \beta & = & \alpha (3-\tau_c). \label{eq:4}
 \eea
 The model when $\mu=0$ was studied using a mean field approximation
that ignored the spatial correlations between masses at different sites
\cite{inout2}. It was shown that $\phi=1$, $\alpha=2/3$ and $\tau_c=5/2$
when $\mu=0$. Correspondingly, $\theta = 2$ and $\beta = 1/3$.  In one
dimension, the exponents when $\mu=0$ were determined numerically to be
$\tau_c \approx 1.83$, $\alpha \approx 0.61$, $\phi \approx 1.01$,
$\theta \approx 1.47$, and $\beta \approx 0.71$.

\subsection{\label{sec2c} Summary of results}

In this paper, the In-Out model is studied for $\mu \geq 0$. Using
scaling arguments, a relation is derived between the exponents $\alpha$
and $\tau_c$, thus reducing the number of unknown exponents from three
to two.  In particular, it is shown that
 \be
 \alpha ( \mu d + 2 \tau_c - 2) = d, \quad d\leq 2.
 \ee

The exponent $\tau_c$ in $d\leq 2$ is shown to be
 \be
 \tau_c=\cases{
 \frac{d^2+6 d + 4}{2 (d+2)} &$\mu =0$, \cr
 \frac{(3-\mu)d +2}{d+2} &$ 0<\mu<2$. \cr}
 \ee
 The exponent $\alpha$ in $d\leq 2$ is shown to be
 \be
 \alpha=\cases{
 \frac{d+2}{d+4} &$\mu =0$, \cr
 \frac{d+2}{\mu d + 4} &$0<\mu<2$. \cr}
 \ee
 The exponent $\phi$ in $d\leq 2$ is calculated for $\mu=0$ and $\mu=1$:
 \be
 \phi=\cases{
 1 &$\mu =0$, \cr
 \frac{d+2}{2} &$\mu=1$. \cr}
 \ee
 In the growing phase, the exponent $\tau$ characterizing the power law
decay of the mass distribution is shown to be
 \be
 \tau = \frac {(2-\mu)d +2}{d+2}.
 \ee

In one dimension, when $\mu=0$, the exponents reduce to $\tau_c=11/6$,
$\alpha=3/5$ and $\phi=1$. This is in very good agreement with the
numerical results seen in Ref.~\cite{inout2} (see Sec.~\ref{sec2b}). In
dimensions greater than two, the exponents take on their mean field
value, obtained by setting $d$ to $2$ in the above equations.

\section{\label{sec3} Scaling relation between $\alpha$ and $\tau_c$}

In this section, a relation between $\alpha$ and $\tau_c$ is obtained
from scaling arguments for all $\mu$. The dependence of the largest mass
in the system $M_t$ on $t$ can be obtained by the catchment area
argument as follows. Due to diffusion, the mass $M_t$ would sweep out an
area $L_t^d$ in time $t$, where $L_t$ is the typical length scale in the
system. In addition to the mass contained in this area, $M_t$ also
increases due to the average flux $F=d \langle m \rangle /dt$. Thus,
 \be
 M_t \sim L_t^d F t. 
 \label{eq:5}
 \ee
 The typical length $L_t$ arises from diffusion: $L_t \sim (D t)^{1/2}$.
 Substituting $M_t^{-\mu}$ for the diffusion rate $D$, and using $F \sim
t^{\alpha (2-\tau_c)-1}$ from the scaling relation Eq.~(\ref{eq:2}), we
obtain $M_t \sim t^{(d+2\alpha (2-\tau_c))/(2+\mu d)}$. But, by
definition (see Eq.~(\ref{eq:2})) $M_t \sim t^{\alpha}$. Equating the
exponents, we obtain
 \be
 \alpha ( \mu d + 2 \tau_c - 2) = d. 
 \label{eq:6}
 \ee
 Thus, the number of unknown exponents reduces from three to two. The
above scaling arguments are valid only when $L_t$ increases as a
positive power of $t$. This restriction translates to the condition
$\alpha \mu <1$.
 
\section{\label{sec4} The growing phase ($q>q_c$)}

In this section, the behavior of $P(m)$ in the growing phase $q>q_c$ is
discussed. The exponent $\tau$ (as defined in Eq.~(\ref{eq:1})) is
expected to be independent of the precise values of $q$ and $p$ as long
as we are above the critical threshold \cite{inout2}. To obtain $\tau$,
the convenient limit $p=0$ and $q$ arbitrary may be studied. Different
aspects of this limiting case have been studied in the context of river
networks, self organized criticality and epitaxial growth
\cite{takayasu, RM1,maritan,KMR,camacho,colm}. We give a short
derivation of the exponent $\tau$. In this limiting case of only
adsorption, it is known that $P(m,t)$ has the scaling form
 \be
 P(m,t)\sim m^{-\tau} f\left(\frac{m}{t^{\delta}}\right),
 \label{eq:7}
 \ee
 where the scaling function $f(x)$ tends to a constant (for $\mu<1$) for
small values of $x$ and decays exponentially for large values of $x$. 
Since there is a constant influx $F$ of particles, $\langle m \rangle =
F t$. Therefore $\delta(2-\tau)=1$.  To obtain a second relation between
the exponents, note that Eq.~(\ref{eq:6}) is valid when $\tau_c$ is
replaced by $\tau$ and $\alpha$ by $\delta$. Solving these two exponent
equalities, one obtains
 \bea
 \tau& =& \frac {(2-\mu)d +2}{d+2}, \nonumber\\
 \delta &=& \frac{d+2}{\mu d +2}. 
 \label{eq:8}
 \eea
 Equation~(\ref{eq:8}) is valid when $\mu<1$ and $d\leq 2$. For $d>2$,
the mean field results are correct. The $\mu=0$, results were obtained
earlier in \cite{takayasu,RM1,maritan}. For $\mu>0$, the one and two
dimensional results were obtained earlier \cite{KMR,camacho}.  The
dependence of $P(m,t)$ on the flux $F$ can now be incorporated into
Eq.~(\ref{eq:7}) by simple dimensional arguments: 
 \be
 P(m,t)\sim \frac{F^{\frac{d}{d+2}}}{m^{\tau}}
f\left(\frac{m}{(F^{\frac{2}{d+2}} t)^{\delta}}\right),
 \label{eq:9}
 \ee
 where $\tau$ and $\delta$ are as in Eq.~(\ref{eq:8}). 

The two variable scaling function $Y(x,y)$ in Eq.~(\ref{eq:2}) should be
such that when $x\gg 1$ (or $m\tilde{q}^\phi \gg 1$), it reduces to the
one variable scaling function $f$ in Eq.~(\ref{eq:9}). This implies that
$Y(x,y) \sim x^{\tau_c-\tau} f(y/x^{1-\alpha/\delta})$ when $x \gg 1$. 
Thus
 \be
 P(m,t)\sim \frac{\tilde{q}^{\phi(\tau_c-\tau)}}{m^{\tau}}
f\left(\frac{m}{(\tilde{q}^\gamma t)^{\delta}}\right),
 \label{eq:10}
 \ee
 where $\gamma$ is a crossover exponent. To make a comparison with
Eq.~(\ref{eq:9}), one has to make the identification $F \sim
\tilde{q}^{\theta}$. Using Eqs.~(\ref{eq:3}), (\ref{eq:6}) and
(\ref{eq:8}), it is easy to show that $\tilde{q}^{\phi(\tau_c-\tau)}
\sim F^{d/(d+2)}$. Also, 
 \be
 \gamma= 2 \theta /(d+2).
 \label{eq:gamma}
 \ee

\section{\label{sec5} Solvable limits}

In this section, we examine limiting cases of the model which are
analytically tractable. For the sake of continuity of argument, the
details of the calculation are deferred to the appendices.

\subsection{\label{sec5a} Solution for $\mu=1$}

The special case when a mass $m$ diffuses as $m^{-1}$ can be solved by
examining the time evolution of the two point correlations. When
$\mu=1$, certain simplifications occur. We refer to
Appendix~\ref{appendix1} for details. It is shown that the critical
$q_c$ at which the mean mass increases with time is
 \be
 q_c = d p^2 g(p), 
 \ee
 where
 \be
 g(p) = \int_{0}^{2 \pi} \frac{d k_1}{2 \pi} \ldots \int_{0}^{2 \pi}
 \frac{d k_d}{2 \pi} \frac{1}{(1+p) d - \sum_{i=1}^{d} \cos(k_i)}.
 \ee
 The exponents for $\mu=1$ is shown to be (see Appendix~\ref{appendix1})

 \bea
 \tau_c&=& \frac{2 d + 2}{d+2}, \label{eq:11} \\
 \alpha&=& \frac{d+2}{d+4}, \label{eq:12} \\
 \phi&=& \frac{d+2}{2}. \label{eq:13}
 \eea
 Solving for $\theta$ from Eq.~(\ref{eq:3}), we obtain $\theta=(d+2)/2$.

\subsection{\label{sec5b} Mean field theory}

The exponents may be computed in large dimensions by a mean field
analysis. This approximation involves ignoring the correlations between
masses at two different sites. The details of the calculation are
presented in Appendix~\ref{appendix2}. The results are
 \be
 \tau_c=\cases{
 \frac{5}{2} &$\mu =0$, \cr
 2-\frac{\mu}{2} &$ 0<\mu<2$. \cr}
 \label{eq:14}
 \ee
 The exponent $\alpha$ in the mean field equals
 \be
 \alpha=\cases{
 \frac{2}{3} &$\mu =0$, \cr
 \frac{2}{2+\mu} &$0<\mu<2$. \cr}
 \label{eq:15}
 \ee
 The exponent $\phi$ could be computed only for $\mu=0$ and $\mu=1$: 
 \be
 \phi=\cases{
 1 &$\mu =0$, \cr
 2 &$\mu=1$. \cr}
 \label{eq:16}
 \ee
 Using Eqs.~(\ref{eq:3}) and (\ref{eq:gamma}), we obtain
 \bea
 \theta &=& 2 \quad \mu=0,1,\\
 \gamma &=& 1 \quad \mu=0,1. \label{eq:gamma2} 
 \eea
 Comparison with the exact solution for $\mu=1$ or calculating the
dimension $d$ when the mean field exponents satisfy the scaling relation
Eq.~(\ref{eq:6}) shows that the upper critical dimension of the system
is $2$.

\subsection{\label{sec5c} Solution for $d=0$}

In zero dimensions, the problem may be solved for in a straightforward
manner (see Appendix~\ref{appendix3}).  The exponents are independent of
$\mu$, since there is no diffusion. In this case,
 \bea
 \tau_c &=&1, \label{eq:17} \\
 \alpha &=&\frac{1}{2}, \label{eq:18}\\
 \phi &=& 1, \label{eq:19}
 \label{eq:exp_d0}
 \eea
 for all values of $\mu$.  Using Eqs.~(\ref{eq:3}) and (\ref{eq:gamma}),
we obtain
 \bea
 \theta &=& 1,\\
 \gamma &=& 1. \label{eq:gamma0}
 \eea

\section{\label{sec6} Exponents for arbitrary $\mu$ and $d$}

The question remains as to what the values of the exponents are in
arbitrary dimensions. They can be determined with the help of an
assumption. We make the assumption that the critical exponents for a
given $\mu$ are a monotonic function of dimension $d$. This assumption
is reasonable as known exponents for most systems at their critical
point (for example, the Ising model) have this property. 

Consider first the exponents when $\mu=0$.  Notice that the exponent
$\phi$ for $\mu=0$ takes on the same value in the mean field or $d=2$
(see Eq.~(\ref{eq:16})) as well as in $d=0$ (see Eq.~(\ref{eq:exp_d0})).
Assuming that $\phi(d)$ should be monotonic in $d$, we obtain
 \be
 \phi = 1 , \qquad \mu = 0.
 \label{eq:20}
 \ee
 Consider now the exponent $\gamma$ (as defined in Eq.~(\ref{eq:10})).
It takes the value $1$ in $d=0$ (see Eq.~\ref{eq:gamma0}) and in the
mean field limit (see Eq.~\ref{eq:gamma2}). Hence, $\gamma = 1$ in all
dimensions or 
 \be
 \theta = \frac{d+2}{2} , \qquad \mu = 0.
 \label{eq:21}
 \ee

Numerical simulations of the $\mu=0$ model in one dimension had $\phi
\approx 1.01$ and $\theta \approx 1.47$ \cite{inout2} consistent with
the above results. Knowing $\theta$, the exponents $\tau_c$ and $\alpha$
can be solved for from Eqs.~(\ref{eq:3}) and (\ref{eq:6}) to yield
 \bea
 \tau_c & = & \frac{d^2+6d+4}{2 (d+2)} , \quad \mu=0, \label{eq:22}\\
 \alpha & = & \frac{d+2}{d+4}, \quad \mu=0. \label{eq:23}
 \eea
 Specializing to $d=1$, the exponents reduce to $\tau_c=11/6$ and
$\alpha=3/5$. Again, these values are very close to the numerical values
of $1.83$ and $0.61$ obtained in Ref.~\cite{inout2}.
 \begin{figure}
 \includegraphics[width=\columnwidth]{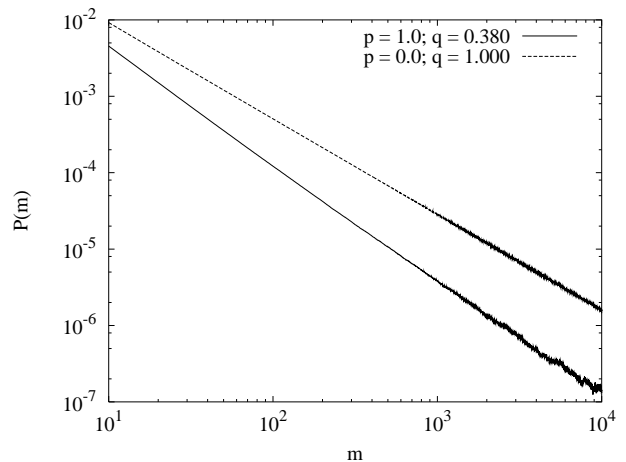}
 \caption{\label{fig1} The variation of $P(m)$ with $m$ is shown at the
critical point (bottom curve) and in the growing phase (top curve). The
results are for $\mu=0.25$. A best fit for the curves give exponent
values $\tau_c=1.58\pm0.04$ and $\tau=1.25\pm 0.01$.}
 \end{figure}

When $\mu >0$, we can calculate the exponent $\tau_c$ and $\alpha$ as
follows. Consider the exponent $\beta$ defined by $\langle m^2 \rangle
\sim t^{\beta}$ at the critical point. Clearly $\beta = \alpha
(3-\tau_c)$. Note that when $\mu>0$, $\beta=1$ in the mean field
analysis as well as in zero dimensions. Thus, using the argument of
monotonicity of exponents, we obtain
 \be
 \alpha (3-\tau_c) = 1, \quad \mu>0.
 \label{eq:24}
 \ee
 Solving Eqs.~(\ref{eq:6}) and (\ref{eq:24}), we obtain 
 \bea
 \tau_c &=& \frac{(3-\mu) d +2}{d+2}, \quad 0<\mu < 2, \label{eq:25}\\
 \alpha &=& \frac{d+2}{\mu d+4}, \quad 0<\mu < 2. \label{eq:26}
 \eea
 When $\mu=1$ or when $d=2$, the results match the exact results derived
in Sec.~\ref{sec4}. The exponent $\phi$ for $\mu>0$ is still
undetermined, and there seems to be no easy way to calculate it.

The analytical results are now compared with results from monte carlo
simulations in one dimension. When $\mu=0$, simulations were done in
Ref.~\cite{inout2}. As pointed out earlier in this section, the
numerical values of the exponents are in close agreement with that
obtained in this paper. We therefore concentrate on non-zero values of
$\mu$. The exponent that is determined numerically is $\tau_c$ for
$\mu=0.25$ and $\mu=0.5$. The simulations were done on a one dimensional
lattice of size $1000$ with periodic boundary conditions. $P(m)$ was
obtained by averaging over $10^8$ realizations. 

In Fig.~\ref{fig1}, the results for $\mu=0.25$ is shown. $P(m)$ is
measured for $q=q_c\approx 0.380$ when $p=1.0$, and for the growing
phase in which $p$ is set to zero. The critical $q_c$ was fixed to be
that value of $q$ at which the distribution changed from an exponential
to a power law. A best fit gives $\tau_c=1.58\pm 0.04$ and
$\tau=1.25\pm0.01$. These should be compared with the analytical results
$\tau_c=1.583\ldots$ and $\tau = 1.25$.

\begin{figure}
 \includegraphics[width=\columnwidth]{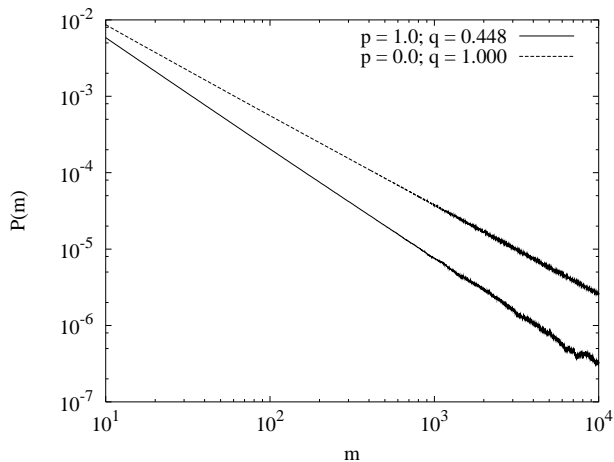}
 \caption{\label{fig2} The variation of $P(m)$ with $m$ is shown at the
critical point (bottom curve) and in the growing phase (top curve). The
results are for $\mu=0.50$. A best fit for the curves give exponent
values $\tau_c=1.47\pm0.04$ and $\tau=1.17\pm 0.01$.}
 \end{figure}

Fig.~\ref{fig2} is as in Fig.~\ref{fig1}, except that $\mu=0.5$ and
$q_c\approx 0.448$. A best fit gives $\tau_c=1.47\pm 0.04$ and
$\tau=1.17\pm0.01$. These should be compared with the analytical results
$\tau_c=1.5\ldots$ and $\tau = 1.166\ldots$.

\section{\label{sec7} Connection to related models}

In this section, similarities between the In-Out model and other related
models of aggregation are discussed.  A model that closely resembles the
model studied in this paper is the charge model with adsorption
\cite{takayasu1,nagatani,MS}. In this model, there is no longer the
restriction that the masses have to be non-negative. Also, $+1$ and $-1$
masses are input at the same rate. In the limit of large time, $P(m)$
for this model is a power law $P(m) \sim |m|^{-5/3}$ for $|m| \gg 1$ in
one dimension.  This model could be expected to have the same behavior
as the In-Out model at the critical point, since the growth velocity is
zero. However, the exponent for the charge model is different from the
value $11/6$ obtained for the In-Out model, showing that the restriction
of non-negative masses is relevant. We now ask whether it is relevant
when $\mu>0$.

Since the charge model was not studied earlier for $\mu>0$, we now give
a short derivation of the power law exponent using scaling arguments. It
was shown based on very general arguments that for the charge model,
irrespective of the diffusion rates that \cite{MS}
 \be
 \langle m^2 \rangle \sim t, \quad t\gg 1.
 \ee
 Assuming scaling for $P(m,t)$ as in Eq.~(\ref{eq:7}) and the scaling
relation Eq.~(\ref{eq:6}), it is easy to see that $P(m)\sim
m^{-\tau_{ch}}$, where 
 \be
 \tau_{ch} = \frac{(3-\mu) d +2}{d+2}, \quad 0<\mu < 2\\
 \ee
 But this is the same as $\tau_c$ obtained for the In-Out model (see
Eq.~(\ref{eq:25}). Thus, the charge model and the In-Out model appears
to have the same behavior when $\mu>0$.

A reason for this could be the following. When $\mu=0$, there is a
chance that large positive masses get neutralized by large negative
masses in the charge model. This process is totally absent in the In-Out
model, resulting in the exponents being different. When $\mu>0$, large
positive and negative charges get immobilized and their collision
becomes infrequent. Hence, one could ignore this process in the charge
model and hence the two models become qualitatively similar. A pitfall
of this argument is that it predicts that $\tau_c$ for $\mu=0$ should be
less than $\tau_{ch}$, contrary to what is seen. Thus, the exact
connection between these two models remains unclear.

Another model which is related to the In-Out model is a model of
coagulation with fragmentation \cite{inout1,mkb1,mkb2}.  In this model,
the desorption at a site is accompanied by adsorption at the neighboring
site, thus conserving mass locally. In this model, there is a phase
transition from a phase in which $P(m)$ is exponentially distributed to
one in which it is a power law, accompanied by a infinite aggregate
which accommodates a finite fraction of the total mass. When the
diffusion constant is mass-dependent, it was shown that \cite{mkb2}, the
exponent $\tau_c$ in the mean field limit is exactly the same as that of
the In-Out model in the mean field limit. However, in dimensions lower
than the upper critical dimension, the exponents in the model with mass
conservation remains equal to the mean field value, unlike the In-Out
model.

\section{\label{sec8} Summary and conclusions}

To summarize, the exponents characterizing the phase transition from a
phase with finite mean height to one in which it grows linearly with
time in the In-Out model were calculated. The model was extended to one
in which particles diffuse with a mass dependent rate $D(m) \propto
m^{-\mu}$. The exponents were shown to have a discontinues jump at
$\mu=0$. The exponents are unrelated to previously studied universality
classes of nonequilibrium phase transitions.

There are several questions that remain unanswered. Other models which
show a wetting transition as seen in the In-Out model have exponents
which can be expressed in terms of exponents of absorbing phase
transitions \cite{alon,politi}.  Here, there seems to be no apparent
connection to any underlying absorbing phase transition.  It would be
interesting to find connections to other models of nonequilibrium phase
transitions.

The calculation of exponents in this paper for arbitrary $\mu$ relied on
the assumption that the exponents are monotonic with dimension. While
simulations do support the results that are obtained, it is important to
have a more rigorous derivation of the exponents without making this
assumption. Also, one would expect logarithmic corrections to the power
laws in two dimensions. These have been ignored in this paper. A
calculation of these corrections would be of interest.

A connection to the charge model was pointed out in Sec.~\ref{sec7}. The
models seem to be similar for $\mu>0$, while different for $\mu=0$. The
precise connection between the two would be worth exploring since the
charge model is analytically more tractable. Finally, the discontinuity
of exponents at $\mu=0$ remains a puzzle.

\appendix

\section{\label{appendix1} Exact solution for $\mu=1$}

In this appendix, we derive the exponents for $\mu=1$ in arbitrary
dimensions. We do so by examining the two point correlations in the
system in the steady state. To fix notation, let ${\bf x}'$ denote one
of the $2d$ nearest neighbors of the site ${\bf x}$. Let $\eta({\bf
x},{\bf x}',t)$ be the mass transferred from site ${\bf x}$ to ${\bf
x}'$ at time $t$ in a time interval $\Delta t$. From the definition of
the model, it follows that
 \be
 \eta({\bf x},{\bf x}',t) = \cases{
 m_{{\bf x}} & with prob. $\frac{1}{2 d}\frac{\Delta t}{m_{{\bf
x}}^{\mu}}$, \cr
 0 & otherwise. \cr}
 \label{eq:a1} 
 \ee
 To order $\Delta t$, the only nonzero two point correlation in the
noise is
 \be
 \langle \eta({\bf x},{\bf x}^\prime)^2\rangle = \frac{m_{{\bf
x}}^{2-\mu} \Delta t}{2 d}.
 \label{eq:a2}
 \ee
 Let $I({\bf x},t)$ be the mass transferred due to adsorption and
desorption from the site ${\bf x}$ at time $t$ in an infinitesimal time
$\Delta t$. Then,
 \be
 I({\bf x},t) = \cases{
 1 & with prob $q \Delta t$, \cr
 -1+\delta_{m_{{\bf x}},0} & with prob $p \Delta t$, \cr
 0 & otherwise. \cr}
 \label{eq:a3} 
 \ee
 To order $\Delta t$, the only nonzero two point correlation in the
input $I$ is
 \be
 \langle I({\bf x},t)^2\rangle = (q+p s) \Delta t,
 \label{eq:a4}
 \ee
 where $s=\sum_{m=1} P(m)$ is the occupation probability.

The mass $m_{{\bf x}}(t)$ at lattice site ${\bf x}$ at time $t$ evolves
as
 \bea
 m_{{\bf x}}(t+{\Delta t})& = & m_{{\bf x}}(t) - \sum_{{\bf x}'}
\eta({\bf x},{\bf x}',t) + \sum_{{\bf x}'}\eta({\bf x}',{\bf x},t)
\nonumber\\ &&\mbox{} + I({\bf x},t).
 \label{eq:a5}
 \eea
 To obtain the two point correlations, we multiply $m_{{\bf
x}}(t+{\Delta t})$ by $m_{{\bf 0}}(t+{\Delta t})$ and take averages over
the possible stochastic moves.  Using Eqs.~(\ref{eq:a1})--(\ref{eq:a5}),
we obtain \begin{widetext}
 \bea
 \frac{d C( {\bf x})}{dt} &=& - 2 C_{\mu} ({\bf x})+ \frac{1}{d}
\sum_{j=1}^d \sum_{k=\pm 1} C_{\mu} (x_1, \ldots, x_j + k, \ldots x_d) +
2 (q-p) \rho + 2 p D({\bf x}) \nonumber \\ &&+ \delta_{{\bf x},0} (q+p
s) + \langle m^{2 - \mu} \rangle(2 \delta_{{\bf x},{\bf 0}} -
\frac{1}{d} \sum_{j=1}^{d} \sum_{k=\pm 1} \delta_{x_1,0} \ldots
\delta_{x_j+k,0} \ldots \delta_{x_d,0} \rangle
 \label{eq:a6}
 \eea 
 \end{widetext}
 where $C({\bf x})=\langle m_{{\bf x}}m_{{\bf 0}} \rangle$,
$C_{\mu}({\bf x})=\langle m_{{\bf x}}m_{{\bf 0}}^{1-\mu} \rangle$,
$\rho=\langle m \rangle$ and $D({\bf x}) = \langle m_{{\bf x}}
\delta_{m_{{\bf 0}},0} \rangle$. 

Consider Eq.~(\ref{eq:a6}) in the steady state when the time derivative
is set to zero.  For arbitrary values of $\mu$, the right hand side of
Eq.~(\ref{eq:a6}) involves three unknowns: $C_{\mu}({\bf x})$, $D({\bf
x})$ and $\rho$. However, when $\mu=1$, a simplification occurs because
 \be
 C_{1}({\bf x}) = \rho - D({\bf x}),
 \label{eq:a7}
 \ee
 thus reducing the number of unknowns to two. Define
 \be
 F({\bf k}) = \sum_{{\bf x}} \left[ D({\bf x}) - \rho(1-s) \right] e^{i
{\bf k}. {\bf x}}.
 \label{eq:a8}
 \ee
 Solving for $F({\bf k})$ from Eq.~(\ref{eq:a6}), we obtain
 \be
 F({\bf k}) = -\rho + \frac{q d - p \rho d}{h ({\bf k})},
 \label{eq:a9}
 \ee
 where
 \be
 h ({\bf k}) = \sum_{i=1}^{d} \cos(k_i) - (1+p) d. 
 \label{eq:a10}
 \ee
 To obtain $\rho$, we use the fact that the constant term in $F({\bf
k})$ equals $-\rho(1-s)$. Then
 \be
 \rho = \frac{ d p q g(p)}{d p^2 g(p) - q},
 \label{eq:a11}
 \ee
 where
 \be
 g(p) = \int_{0}^{2 \pi} \frac{d k_1}{2 \pi} \ldots \int_{0}^{2 \pi}
\frac{d k_d}{2 \pi} \frac{1}{(1+p) d - \sum_{i=1}^{d} \cos(k_i)}.
 \label{eq:a12}
 \ee
 Thus, the mean density diverges as $(q-q_c)^{-1}$, where
 \be
 q_c = d p^2 g(p). 
 \label{eq:a13}
 \ee
 Specializing the result to one and two dimensions,
 \bea
 q_c^{1D} &=& \frac{p^{3/2}}{\sqrt{p+2}}, \nonumber\\ q_c^{2D} &=&
\frac{2 p^{2}}{\pi (1+p)} K \left[ \frac{1}{1+p} \right],
 \label{eq:a14}
 \eea
 where $K$ is the complete elliptic integral of the first kind. 

 The form of the divergence of the density as $q$ approaches $q_c$,
namely $\langle m \rangle \sim \tilde{q}^{-1}$ means that
 \be
 \phi(2 - \tau_c) = 1, \quad \mbox{for}~\mu=1,
 \label{eq:a15}
 \ee
 in all dimensions. To obtain one more relation between the exponents,
we consider $\frac{d \langle m^2 \rangle}{dt}$ for large times at the
critical point. Firstly, we need to invert Eq.~(\ref{eq:a9}) to obtain
$D({\bf 1})$ where ${\bf 1}$ denotes the site $(1,0,0,\ldots)$.
Inverting, we obtain
 \be
 D({\bf 1}) = \frac{\rho(p-q)}{q} + (q-p \rho) \left[1 - d (1+p) g(p)
\right].
 \label{eq:a16}
 \ee
 We now make the assumption that the leading time dependence to $D({\bf
1})$ is obtained by restoring the time dependence of $\rho$. Then, when
${\bf x}={\bf 0}$, Eq.~(\ref{eq:a6}) reduces to
 \be
 \frac{d \langle m^2 \rangle}{dt} \approx 2 \rho(t)
\frac{(1+p)(q-q_c)}{p} +2 q d (1+p) g(p).
 \label{eq:a17}
 \ee 
 If we now take the limit $q\rightarrow q_c$ before $t \rightarrow
\infty$, then we obtain that $\langle m^2 \rangle \sim t$. Thus
 \be
 \alpha(3-\tau_c) = 1, \quad \mbox{for}~\mu=1,
 \label{eq:a18}
 \ee
 in all dimensions. Solving for the exponents $\tau_c$, $\alpha$ and
$\phi$ from Eqs.~(\ref{eq:6}),(\ref{eq:a15}) and (\ref{eq:a18}), we
obtain
 \bea
 \tau_c &=& \frac{2 d + 2}{d+2}, \label{eq:a19} \\
 \alpha &=& \frac{d+2}{d+4}, \label{eq:a20}\\
 \phi &=& \frac{d+2}{2}.\label{eq:a21}
 \eea
 Correspondingly $\theta = (d+2)/2$ and $\beta=1$.

\section{\label{appendix2} Mean field solution}

In this appendix, we derive the exponents for $\mu\geq 0$ using a mean
field approximation. This approximation involves ignoring correlations
between the masses, i.e., replacing joint probability distribution
functions by product of single point distributions. Then, the master
equation for the temporal evolution of $P(m,t)$ is \bea
 \frac{d P(m)}{d t} &=& -(m^{-\mu}+s'+p+q) P(m) + q P(m-1)\nonumber \\ &
+& p P(m+1) + \sum_{m'=1}^{m} \frac{P(m') P(m-m')}{m'^{\mu}}
\label{eq:b1}\\
 \frac{d P(0)}{dt} &=& s s' - q + q s + p P(1) \label{eq:b2},
 \eea
 where $s=\sum_{m=1} P(m)$ and $s'= \sum_{m=1} m^{-\mu} P(m)$. The
different terms enumerate the number of ways the mass at a certain site
can change. Then it follows that \bea
 \frac{d \langle m^n \rangle}{dt}& = &\sum_{k=1}^{n-1} {n \choose k}
\langle m^k \rangle \langle m^{n-k-\mu} \rangle + q + (-1)^n p
s\nonumber \\ &+& \sum_{k=1}^{n-1} {n \choose k} [q + (-1)^k p] \langle
m^{n-k} \rangle.
 \label{eq:b3}
 \eea
 Consider first the steady state solution of Eq.~(\ref{eq:b3}) when the
time derivatives may be set to zero. Then, putting $n=1$, and solving
for the occupation probability $s$, we obtain
 \be
 s=\frac{q}{p}.
 \label{eq:b4}
 \ee

The mean field equations take on a simpler form for the cases $\mu=0$
and $\mu=1$, and hence we solve them separately from the arbitrary $\mu$
case.  Though all the mean field exponents for $\mu=0$ were derived in
Ref.~\cite{inout2} using the generating function method, they will be
rederived here using a different method which will be simpler to
generalize to the $\mu \geq 0$ case. 

\subsection{\label{appendix2a} $\mu=0$}

On choosing $n=2$ in Eq.~(\ref{eq:b3}) and taking the steady state
limit, a quadratic equation for $\langle m \rangle$ is obtained which
can be solved to yield
 \be
 \langle m \rangle = \frac{ p - q - \sqrt{(p-q)^2 - 4 q}}{2}
 \label{eq:b5}
 \ee
 The expression for $\langle m \rangle$ becomes invalid when expression
under the square root sign becomes negative, thus fixing $q_c$. 
Solving, we obtain
 \be
 q_c = p+2 - 2 \sqrt{1+p},
 \label{eq:b6}
 \ee
 where the sign is chosen by the condition that $q_c=0$ at $p=0$.
Consider now the equation corresponding to $n=3$ in Eq.~(\ref{eq:b3}).
Solving for $\langle m^2 \rangle$, we obtain
 \be
 \langle m^2 \rangle = \frac{ (p+q) \langle m \rangle} {p-q - 2 \langle
m \rangle}.
 \label{eq:b7}
 \ee
 But the denominator tends to zero as $\sqrt{\tilde{q}}$, where
$\tilde{q}= q_c-q$. Therefore, near the transition point $\langle m^2
\rangle$ diverges as
 \be
 \langle m^2 \rangle \sim \frac{1}{\tilde{q}^{1/2}}.
 \label{eq:b8}
 \ee
 Consider now the equation corresponding to $n=4$ in Eq.~(\ref{eq:b3}).
Solving for $\langle m^3 \rangle$, we obtain
 \bea
 \langle m^3 \rangle &=&\! \frac{3 \langle m^2 \rangle^2\! +\! 3 (q+p) \langle
m^2 \rangle \!+ \!2 (q-p) \langle m \rangle\!+\!q } {2 (p-q - 2 \langle m
\rangle)} \label{eq:b9}\\ 
 &\sim & \frac{1}{\tilde{q}^{3/2}}, \quad \tilde{q} \rightarrow 0.
 \label{eq:b10}
 \eea

Knowing the behavior of $\langle m^2 \rangle$ and $\langle m^3 \rangle$
near the transition point, we immediately obtain
 \bea
 \phi(3-\tau_c)& =& \frac{1}{2}, \label{eq:b11}\\
 \phi(3-\tau_c)& =& \frac{3}{2}, \label{eq:b12}
 \eea
 which can be solved to give
 \bea
 \tau_c &=& 5/2, \label{eq:b13}\\
 \phi &=& 1. \label{eq:b14}
 \eea

To calculate $\phi$ and $\tau$ we had first taken the limit $t
\rightarrow \infty$ followed by the limit $\tilde{q} \rightarrow 0$. In
order to calculate $\alpha$, we need to take the limits in the opposite
order, namely $\tilde{q} \rightarrow 0$ followed by $t \rightarrow
\infty$.  We first note that in this limit $s = q/p$ and $\langle m
\rangle = (p-q)/2$. Then choosing $n=3$ in Eq.~\ref{eq:b3}, we obtain 
 \be
 \frac{d \langle m^3 \rangle}{ dt} = \frac{3 (p^2-q_c^2)}{2}.
 \label{eq:b15}
 \ee
 Thus,
 \be
 \alpha (4-\tau_c)=1.
 \label{eq:b16}
 \ee
 Substituting for $\tau_c$, we obtain
 \be
 \alpha = \frac{2}{3}.
 \label{eq:b17}
 \ee
 Thus, in the mean field limit, the scaling function takes on the form
 \be
 P(m,\tilde{q}, t) \sim \frac{1}{m^{5/2}} Y \left( m \tilde{q},
\frac{m}{t^{2/3}} \right), \qquad \mu=0.
 \label{eq:b18}
 \ee

\subsection{\label{appendix2b} $\mu=1$}

We start again with Eqs.~(\ref{eq:b3}) and (\ref{eq:b4}). For $\mu=1$,
Eq.~(\ref{eq:b3}) simplifies because $\langle m^{n-k-\mu} \rangle$
reduces to an integer moment of $m$. Choosing $n=2$ in Eq.~(\ref{eq:b3})
in the steady state, we obtain 
 \be
 \langle m \rangle = \frac{p q}{p^2- p q -q}.
 \label{eq:b19}
 \ee
 The mean mass $\langle m \rangle$ diverges when $q=q_c$, where 
 \be
 q_c = \frac{p^2}{1+p}.
 \label{eq:b20}
 \ee
 Choosing $n=3$ in Eq.~(\ref{eq:b3}) in the steady state, we obtain
 \be
 \langle m^2 \rangle = p\frac{\langle m \rangle^2 + (q+p) \langle m
\rangle}{p^2 - p q -q}.
 \label{eq:b21}
 \ee
 Equations~(\ref{eq:b19}) and (\ref{eq:b21}) imply that $\langle m
\rangle \sim \tilde{q}^{-1}$ and $\langle m^2 \rangle \sim
\tilde{q}^{-3}$ when $\tilde{q} \rightarrow 0$. Thus,
 \bea
 \phi(2-\tau_c) & = & 1, \label{eq:b22}\\
 \phi(3-\tau_c) & = & 3. \label{eq:b23}
 \eea
 Solving, we obtain
 \bea
 \tau_c &=& \frac{3}{2}, \label{eq:b24}\\
 \phi&=& 2. \label{eq:b25}
 \eea

In order to calculate $\alpha$, we need to set $q=q_c$ and take the
large time limit. Then choosing $n=3$ in Eq.~\ref{eq:b3}, we obtain 
 \be
 \frac{d \langle m^2 \rangle}{dt} \sim t^{2 \alpha_c (2-\tau)-1} +
\frac{2 p^2} {(1+p)}.
 \label{eq:b26}
 \ee
 If we assume that the first term is the dominant term, then we reach a
contradiction for $\tau$ (namely, $\tau=1$). The other alternative is to
that $\langle m^2 \rangle \sim t$, implying that $\alpha(3-\tau_c)=1$ or
 \be
 \alpha = \frac{2}{3}.
 \label{eq:b27}
 \ee
 Thus, in the mean field limit, the scaling function takes on the form
 \be
 P(m,\tilde{q}, t) \sim \frac{1}{m^{3/2}} Y \left( m \tilde{q}^2,
\frac{m}{t^{2/3}} \right), \qquad \mu=1.
 \label{eq:b28}
 \ee

\subsection{\label{appendix2c} $\mu>0$}

We will follow the same procedure as for the $\mu =0$ and the $\mu=1$
cases. However, for arbitrary $\mu$, we are no longer able to determine
neither the critical value $q_c$ nor the exponent $\phi$. Consider the
equations arising from choosing $n=2$ and $n=3$ in Eq.~(\ref{eq:b3}) in
the steady state:
 \bea
 \langle m^{1-\mu} \rangle & = & p-q -\frac{q}{\langle m \rangle},
\label{eq:b29}\\
 \langle m^{2-\mu} \rangle & = & \frac{ q \langle m^2 \rangle}{ \langle
 m \rangle^2} - (p+q). \label{eq:b30}
 \eea
 To satisfy Eq.~(\ref{eq:b30}), we require that $\langle m \rangle$
diverges at the critical point. Substituting the scaling form, we obtain
 \be
 \phi(3 - \mu - \tau_c) = \phi(3 - \tau_c) - 2 \phi(2 - \tau_c)
 \label{eq:b31}
 \ee
 implying that
 \be
 \tau_c = 2-\frac{\mu}{2}.
 \label{eq:b32}
 \ee

To calculate the exponent $\alpha$, all we require is that $\langle
m^{1-\mu} \rangle$ is finite at the critical point and is equal to
$p-q$. This follows from Eq.~(\ref{eq:b29}). Now, if we stay at the
transition point, we obtain that $d \langle m^2 \rangle /dt = 2 q$
implying that
 \be
 \langle m^2 \rangle \sim t.
 \label{eq:b33}
 \ee
 Therefore $ \alpha (3-\tau_c) =1$ or
 \be
 \alpha = \frac{2}{2+\mu}.
 \label{eq:b34}
 \ee
 Thus, in the mean field limit, the scaling function takes on the form
 \be
 P(m,\tilde{q}, t) \sim \frac{1}{m^{2-\mu/2}} Y \left( m \tilde{q}^\phi,
\frac{m}{t^{2/(2+\mu)}} \right), \quad 0< \mu<2.
 \label{eq:b35}
 \ee

\section{\label{appendix3} Solution for $d=0$}

In zero dimensions, the problem becomes analytically tractable as
diffusion no longer plays a role. Therefore, the exponents are
independent of $\mu$. The master equation for the evolution of the mass
distribution $P(m)$ is 
 \bea
 \frac{d P(m)}{dt} &=& \!-(p\!+\!q) P(m) \!+ \!p P(m\!+\!1)\! +\! q P(m\!-\!1),
\label{eq:c1} \\
 \frac{d P(0)}{dt}& =& -q P(0) + p P(1). \label{eq:c2}
 \eea
 The steady state solution is obtained by setting the time derivatives
to zero. It is then straightforward to obtain
 \be
 P(m) = \frac{p-q}{p}\left(\frac{q}{p}\right)^m, \quad m\geq 0.
 \label{eq:c3}
 \ee
 This solution is valid when $q<p$. For $q \geq q_c=p$, there is no
nontrivial steady state solution. The typical mass diverges as $q$
approaches $q_c$ as $(q-q_c)^{-1}$; therefore
 \be
 \phi=1.
 \label{eq:c4}
 \ee
 Also, the occupation probability $s=1$ when $q=q_c$. Since $s$ cannot
increase beyond $1$, it remains stuck at $1$ for all further values of
$q$. When $\tilde{q}=q-q_c$ is positive, $d\langle m \rangle/dt = q-p s
= \tilde{q}$. This means that $\theta=1$.

The exponents $\tau_c$ and $\alpha$ may be obtained by solving the
problem at $q=q_c$. In this case if one were to identify $m$ as the
coordinate of a random walker, then the problem reduces to a problem of
a random walker with a reflecting barrier at the origin. This problem is
easily solved \cite{feller} and in the limit of large time, 
 \be
 P(m,t) \approx \frac{1}{\sqrt{\pi q t}} \exp\left(\frac{- m^2}{4 q
t}\right), ~t\gg 1.
 \label{eq:c5}
 \ee
 The exponents $\tau_c$ and $\alpha$ may be read off from
Eq.~(\ref{eq:c5}) to be
 \bea
 \tau_c &=&1, \label{eq:c6} \\
 \alpha &=&\frac{1}{2}, \label{eq:c7}
 \eea
 in zero dimensions.

When $q>q_c$, the problem reduces to the problem of a random walker with
a drift and a reflecting barrier at the origin. Again, this problem is
easily solvable:
 \be
 P(m,t)=\frac{1}{m} \frac{m}{(q-p) t} \delta\left(\frac{m}{(q-p) t} -1
\right),~~m,t \gg 1.
 \label{eq:c8}
 \ee
 Clearly, the exponent $\gamma=1$ (see Eq.~(\ref{eq:10}) for
definition).


\begin{thebibliography}{1}

\bibitem{epitaxy} J.~A.~Venables, G.~D.~T.~Spiller, and
M.~Hanb\"{u}cken, Rep. Prog. Phys. {\bf 47}, 399 (1984).

\bibitem{scheidegger} A.~E.~Scheidegger, Bull. I.A.S.H. {\bf 12}, 15
(1967).

\bibitem{dodds} P.~S.~Dodds and D.~H.~Rothman, Phys. Rev. E {\bf 59},
4865 (1999).

\bibitem{fried} S.~K.~Friedlander, {\it Smoke, Dust and Haze} (Wiley
Interscience, New York, 1977).

\bibitem{white} W.~H.~White, J. Colloid Interface Sci. {\bf 87}, 204
(1982).

\bibitem{ziff} R.~M.~Ziff, J. Stat. Phys. {\bf 23}, 241 (1980).

\bibitem{inout1} S.~N.~Majumdar, S.~Krishnamurthy, and M.~Barma, Phys.
Rev. Lett. {\bf 81}, 3691 (1998).

\bibitem{inout2} S.~N.~Majumdar, S.~Krishnamurthy, and M.~Barma, Phys.
Rev. E {\bf 61}, 6337 (2000).

\bibitem{alon} U.~Alon, M.~R.~Evans, H.~Hinrichsen, and D.~Mukamel,
Phys. Rev. Lett. {\bf 76} 2746 (1996); Phys. Rev. E {\bf 57} 4997
(1998).

\bibitem{politi} H.~Hinrichsen, R.~Livi, D.~Mukamel, and A.~Politi,
Phys. Rev. Lett. {\bf 79}, 2710 (1997).

\bibitem{takayasu} M.~Takayasu and H.~Takayasu in {\it Nonequilibrium
Statistical Mechanics in One Dimension} ed. V.~Privman (Cambridge Univ.
Press, Cambridge, 1997).

\bibitem{maritan} M.~R.~Swift, F.~Colaiori, A.~Flammini, A.~Maritan,
A.~Giacometti and J.~R.~Banavar, Phys. Rev. Lett. {\bf 79}, 3278 (1997).

\bibitem{RM1} R.~Rajesh and S.~N.~Majumdar, Phys. Rev. E {\bf 62}, 3186
(2000). 

\bibitem{KMR}P.~L.~Krapivsky, J.~F.~F.~Mendes, and S.~Redner, Phys. Rev.
B {\bf 59}, 15950 (1999); Eur. Phys. J. B {\bf 4}, 401 (1998).

\bibitem{camacho} J.~Camacho, Phys. Rev. E {\bf 63}, 046112 (2001).

\bibitem{colm} C.~Connaughton, R.~Rajesh and O.~Zaboronski,
cond-mat/0310063.

\bibitem{takayasu1} H.~Takayasu, Phys. Rev. Lett. {\bf 63}, 2563 (1989).

\bibitem{nagatani} T.~Nagatani, J. Phys. A {\bf 26}, L489 (1993).

\bibitem{MS} S.~N.~Majumdar and C.~Sire, Phys. Rev. Lett. {\bf 71}, 3729
(1993).

\bibitem{mkb1} R.~Rajesh and S.~N.~Majumdar, Phys. Rev. E {\bf 63},
036114 (2001). 

\bibitem{mkb2} R.~Rajesh, D.~Das, B.~Chakraborty, and M.~Barma, Phys.
Rev. E {\bf 66}, 056104 (2002).

\bibitem{feller} W.~Feller, {\it An introduction to probability theory
and its applications} (Wiley, New York, 1970).

\end{thebibliography}
\end{document}